\documentclass[preprint]{vgtc}
\preprinttext{\parbox{0.95\textwidth}{\centering\small Accepted at VISxGenAI 2026, a workshop co-located with IEEE VIS 2026.}}

\usepackage{times}

\usepackage{mathptmx}
\usepackage{amsmath}
\usepackage{amssymb}
\usepackage{graphicx}
\usepackage{booktabs}
\usepackage{array}
\usepackage{enumitem}
\usepackage{microtype}

\newcommand{\method}{CanvasObs}
\newcommand{\baseline}{tldraw Agent Starter Kit}

\vgtcinsertpkg

\title{CanvasObs: Structured Spatial Observation for Agent Editing on 2D~Infinite~Canvases}

\author{%
    Jiahao Xu\thanks{e-mail: \href{mailto:jiahao@vt.edu}{jiahao@vt.edu}}\\
    \scriptsize Virginia Tech
    \and Haohui Wang\thanks{e-mail: \href{mailto:haohuiw@vt.edu}{haohuiw@vt.edu}}\\
    \scriptsize Virginia Tech
    \and Rebecca Faust\thanks{e-mail: \href{mailto:rfaust1@tulane.edu}{rfaust1@tulane.edu}}\\
    \scriptsize Tulane University
    \and Dawei Zhou\thanks{e-mail: \href{mailto:zhoud@vt.edu}{zhoud@vt.edu}}\\
    \scriptsize Virginia Tech
    \and Chris North\thanks{e-mail: \href{mailto:north@vt.edu}{north@vt.edu}}\\
    \scriptsize Virginia Tech
}
\hypersetup{
    pdftitle={CanvasObs: Structured Spatial Observation for Agent Editing on 2D Infinite Canvases},
    pdfauthor={Jiahao Xu, Haohui Wang, Rebecca Faust, Dawei Zhou, Chris North}
}

\teaser{
    \centering
    \includegraphics[width=0.9\linewidth]{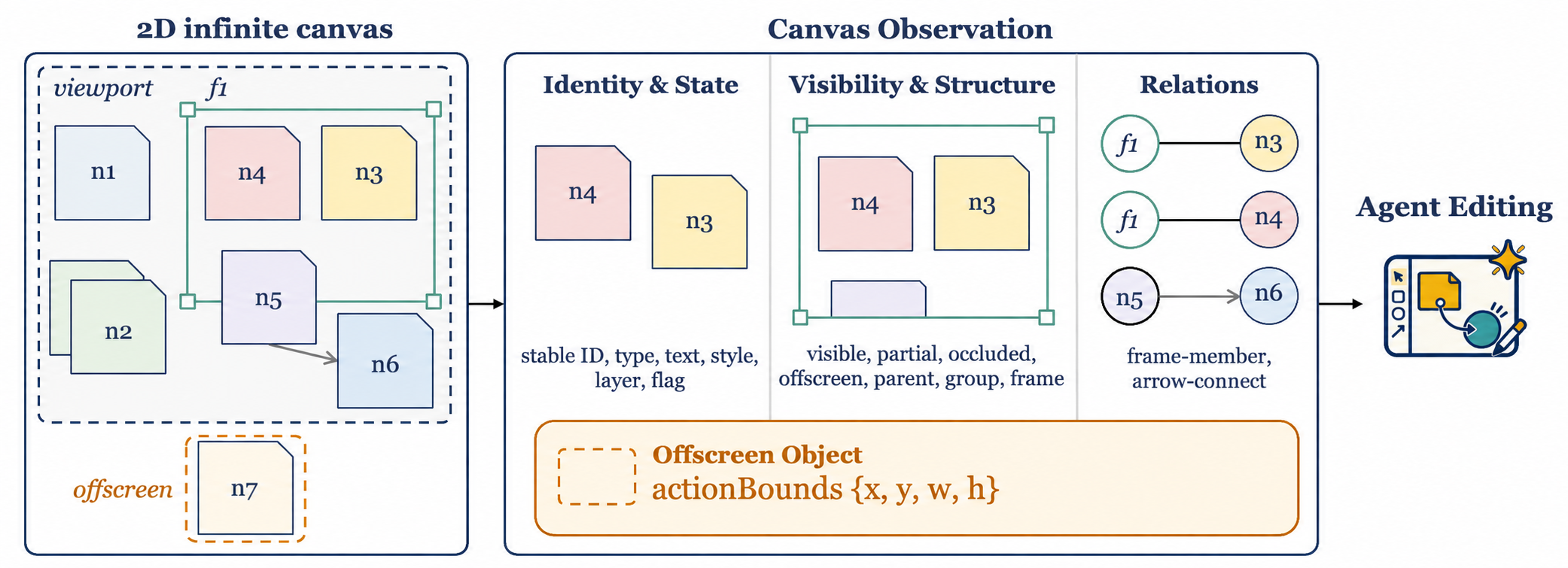}
    \caption{\method{} supplements the shared visual context with a description of objects on the current page, their state and relations, and offscreen bounds in action coordinates. Within each matched pair, the model and editing pipeline remain fixed.}
    \label{fig:teaser}
}

\abstract{
Editing agents on 2D infinite canvases must reason about application state that pixels may not reveal. Targets may be offscreen or covered, while visual containment and touching arrows may disagree with recorded frame membership or connector bindings. We ask how a structured observation layer can support accurate object selection during agentic editing on 2D infinite canvases. We present \method{}, which augments visual context with a structured payload and guidance for reading four components: Identity \& State, Visibility \& Structure, Relations, and Offscreen Object bounds in action coordinates. A Remote Team Handoff case study illustrates retrieving an editable offscreen copy, choosing a covered chart by color, separating cards by frame membership, and preserving an evidence connection. We turn these cases into ten controlled configurations and compare \method{} with the baseline agent in 1,200 browser trials across three models. \method{} improves \emph{Correct target} for every model. Combined across models, it raises \emph{Correct target} from 16.5\% to 48.3\%, an increase of 31.8 percentage points (95\% confidence interval [29.0, 34.7]), and \emph{Full task success} from 4.3\% to 25.0\%. These results show that compact structured observation can improve object choice on 2D infinite canvases without redesigning the editing pipeline.
}


\begin{document}

\firstsection{Introduction}
\maketitle

A \emph{2D infinite canvas} is a pannable, zoomable workspace for arranging editable objects on an effectively unbounded surface~\cite{Suh2023,tldraw2026}. During analysis and sensemaking, people arrange notes, evidence, charts, and intermediate findings to compare alternatives and revisit decisions~\cite{Shipman1995,Wright2006,Andrews2010}. An editing agent must interpret both the visible layout and the underlying document structure, including object identities, layers, parent frames, and connector bindings.

That recovery is difficult when the target lies outside the viewport or beneath another object. Visual containment may disagree with native frame membership, and an arrow that touches a card may not be bound to it. Screenshots communicate appearance, while conventional shape summaries make some objects addressable. Neither necessarily exposes the fact that distinguishes the intended target. If that fact is absent, even a capable planner has no reliable basis for choosing among plausible targets. It may understand the goal yet edit the wrong object.

We therefore ask: \emph{How can a structured observation layer support accurate object selection during agentic editing on 2D infinite canvases?} To investigate this question, we introduce \method{}, a structured observation layer for agent editing on 2D infinite canvases. \method{} supplements the visual context provided by the \baseline{}~\cite{tldraw2026} with four complementary kinds of information, summarized in Figure~\ref{fig:teaser}. 
This description complements rather than replaces the screenshot and existing shape summaries. Within each comparison, both variants use the same model, user prompt, actions, executor, and review lifecycle. \method{} changes only the structured payload and the instructions for interpreting it.

Our evaluation combines the Remote Team Handoff case study (Figure~\ref{fig:case-study}) with a quantitative experiment. Four prompts illustrate retrieving an editable offscreen copy, selecting an occluded chart, separating cards by native frame membership, and preserving an evidence connection. Ten hidden-state configurations, each repeated twenty times with three models and two variants, yield 1,200 browser trials in 600 matched pairs. \method{} improves \emph{Correct target} for every model, but remaining \emph{Correct placement} failures show that better observation does not guarantee a well-executed edit.

This paper makes three contributions:
\begin{itemize}[noitemsep, topsep=0pt, leftmargin=*]
    \item \textbf{\method{}}, a structured observation layer exposing editable object identity, hidden state, relations, and offscreen geometry for agentic editing on 2D infinite canvases.
    \item \textbf{A tldraw implementation} that integrates \method{} while keeping the editing pipeline fixed.
    \item \textbf{A controlled evaluation} demonstrating improved object selection across ten hidden-state configurations.
\end{itemize}

\section{Related Work}

\paragraph{Spatial workspaces and generative AI.}
Spatial layouts help people externalize evolving interpretations. Grouping, alignment, and proximity carry meaning beyond item content~\cite{Shipman1995}, and analytic workspaces use space as external memory and a semantic layer~\cite{Wright2006,Andrews2010}. Generative AI supports chart authoring and editing~\cite{Ye2024,Dibia2023,Wang2024DataFormulator}. Intelligent Canvas and Sensecape embed model assistance in nonlinear workspaces~\cite{Ding2024,Suh2023}, while related systems support collaborative spatial analysis and summarization guided by workspace organization~\cite{Harden2023,Tang2024}. ParaView-MCP combines tool use with visual feedback from a scientific visualization viewport~\cite{Liu2025ParaViewMCP}. \method{} studies editable state absent from the viewport.

\paragraph{Agent observation and element selection.}
Interface agents use screenshots, structured state, or both. Mind2Web and WebArena expose addressable DOM or accessibility elements~\cite{Deng2023,Zhou2024WebArena}, whereas SeeClick and UGround locate elements from images~\cite{Cheng2024,Gou2025UGround}. Hybrid environments combine screenshots with OCR, DOM, or UI automation data~\cite{Koh2024VisualWebArena,Bonatti2025WindowsArena}. OSWorld reports failures in locating and manipulating interface elements, while UI-Vision identifies weaknesses in spatial reasoning and dragging~\cite{Xie2024,Nayak2025UIVision}. \method{} applies hybrid observation to editable spatial documents, where object identity and hidden state guide selection.

\paragraph{Structured agent interfaces.}
WebMCP exposes web functionality as agent-callable tools~\cite{WebMCP2026}, MCP-Web exposes frontend state and actions through MCP~\cite{MCPWeb2026}, and AG-UI supports agent--frontend communication and state synchronization~\cite{AGUI2026}. \method{} focuses on a canvas observation representation combining editable identity, hidden state, native versus inferred relations, and offscreen bounds in action coordinates.

\paragraph{Graphical document structure.}
Rico pairs screenshots with view hierarchies~\cite{Deka2017}, and multimodal UI models combine pixels, text, and hierarchy~\cite{Sunkara2022}. Visualization retrieval incorporates SVG structure~\cite{Li2022Structure}, while diagram understanding represents constituents and relations as parse graphs~\cite{Kembhavi2016}. Visible proxies support offscreen diagram navigation~\cite{Frisch2013}. \method{} connects these ideas to editable object identities and action coordinates.

\section{Method}

\subsection{Conceptual Model}

\method{} complements the workspace with facts stored by the editing application. Figure~\ref{fig:teaser} organizes them into four components.

\paragraph{Identity \& State.}
Identity links each visual mark to a persistent document object, separating items that share the same text or appearance. State describes content, appearance, layer position, and editing status that pixels may not reveal.

\paragraph{Visibility \& Structure.}
Visibility distinguishes objects that are visible, partly visible, covered, or outside the viewport. Structure records document hierarchy so that apparent enclosure is not mistaken for application membership.

\paragraph{Relations.}
Relations capture frame membership and connector binding. Inferred relations describe layout cues such as overlap, alignment, and proximity. Keeping them distinct prevents visual contact from being treated as an authoritative structural connection.

\paragraph{Offscreen Objects.}
An unseen object also needs a spatial extent in the coordinate system used by editing actions. These bounds let the agent move an existing offscreen object instead of recreating it or relying on coarse peripheral cues.

At prompt $t$, let $W_t$ collect the current page's object records, viewport, selection, and task context. \method{} organizes this application state as
\begin{equation}
\mathcal{O}_t = \operatorname{CanvasObs}(W_t)
= \left\langle \mathcal{I}_t,\mathcal{V}_t,\mathcal{R}_t,
\mathcal{B}^{\mathrm{off}}_t \right\rangle,
\label{eq:canvasobs-model}
\end{equation}
The four terms denote Identity \& State, Visibility \& Structure, Relations, and retained Offscreen Object bounds, in that order.

\begin{table}[!t]
    \caption{Canvas observations.}
    \label{tab:observation-comparison}
    \centering
    \normalsize
    \setlength{\tabcolsep}{3pt}
    \renewcommand{\arraystretch}{1.05}
    \begin{tabular}{|>{\raggedright\arraybackslash}p{\dimexpr0.52\columnwidth-2\tabcolsep-2\arrayrulewidth\relax}|>{\centering\arraybackslash}p{\dimexpr0.20\columnwidth-2\tabcolsep-\arrayrulewidth\relax}|>{\centering\arraybackslash}p{\dimexpr0.28\columnwidth-2\tabcolsep-\arrayrulewidth\relax}|}
        \hline
        \textbf{Observation information} & \textbf{Baseline} & \textbf{\method{}} \\
        \hline
        Screenshot & $\checkmark$ & $\checkmark$ \\
        \hline
        Request-region object IDs, text, and bounds & $\checkmark$ & $\checkmark$ \\
        \hline
        Offscreen geometry & Cluster bounds & Cluster + object bounds \\
        \hline
        Offscreen object IDs & --- & $\checkmark$ \\
        \hline
        Style, layer, lock/hidden state & --- & $\checkmark$ \\
        \hline
        Visibility labels & --- & $\checkmark$ \\
        \hline
        Parent, group, and frame IDs & --- & $\checkmark$ \\
        \hline
        Native bindings and inferred spatial relations & --- & $\checkmark$ \\
        \hline
    \end{tabular}
    \par\vspace{-12pt}
\end{table}

\subsection{Implementation}

At each prompt, the implementation instantiates Equation~\ref{eq:canvasobs-model} by traversing current-page objects in layer order. An object's stable identifier, type, abbreviated text, style, rotation, layer index, and selection, context, lock, and hidden flags populate $\mathcal{I}_t$. Its visibility label and hierarchy identifiers populate $\mathcal{V}_t$. Relation construction yields $\mathcal{R}_t$, while eligible offscreen records supply $\mathcal{B}^{\mathrm{off}}_t$.

Objects fully inside the viewport are \texttt{visible}. Intersecting objects are \texttt{partial}. Remaining objects are \texttt{offscreen-near} or \texttt{offscreen-far}. The near threshold is $0.75$ times the larger viewport dimension. A visible or partial object becomes \texttt{occluded} when a higher object that is not hidden, with opacity of at least $0.9$, covers its clipped bounds. This compares bounds rather than pixels. Recorded relation types capture hierarchy and arrow bindings. Inferred types summarize overlap, proximity, alignment, occlusion, and reading order.

Priority proceeds from selected, context, and partial objects to viewport objects, connected objects, and then offscreen objects by distance. The 180-object limit is soft for the first group, while relations have a hard limit of 700. Only retained offscreen records receive \texttt{actionBounds} $\{x,y,w,h\}$ in action coordinates. Here, $x$ and $y$ locate the upper left corner, and $w$ and $h$ give its size. Other records omit these bounds and raw page coordinates.

We implement \method{} using the \baseline{} with tldraw~4.5.8~\cite{tldraw2026}. Table~\ref{tab:observation-comparison} compares observation inputs across conditions. The baseline provides a screenshot, summaries of shapes inside the request bounds, selected identifiers, and anonymous peripheral clusters. Both variants share this context, history, lint feedback, user prompt, model, actions, executor, and review lifecycle. \method{} adds only the serialized Canvas Observation and instructions for reading it, with no new planner or action language.

\begin{figure*}[!t]
    \centering
    \includegraphics[width=0.76\textwidth]{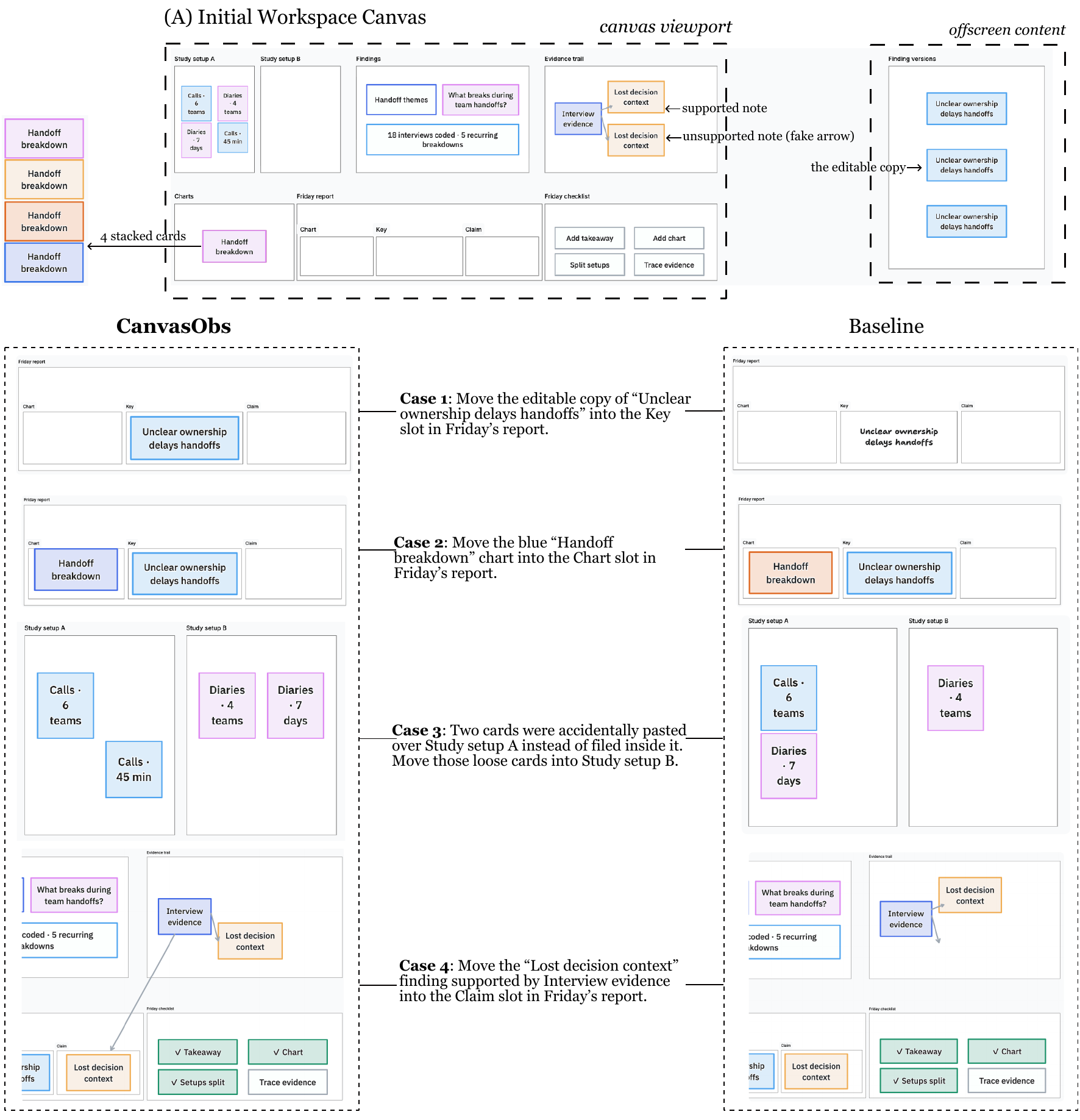}
    \caption{Controlled Remote Team Handoff scenarios. The top panel shows the shared workspace and visible viewport. The version archive lies offscreen. Each lower row pairs a prompt with two example edits. In each row, the left panel shows the intended edit supported by \method{} evidence. The right panel shows a plausible failure when decisive evidence is absent from the baseline input. These panels explain the four ambiguities and are not recorded runs. Section~\ref{sec:quantitative-result} evaluates ten rotated configurations.}
    \label{fig:case-study}
\end{figure*}

\section{Case Study}

Remote Team Handoff is a controlled workflow in which a research team prepares a Friday report from interview evidence, saved findings, study materials, and chart drafts. Four prompts progressively repair the workspace and assemble the report while isolating offscreen identity, occluded state, frame membership, and connector binding. The first two fill the report's Key and Chart slots, the third separates materials for two study setups, and the fourth adds a supported Claim. For each prompt, Figure~\ref{fig:case-study} shows the intended \method{} edit on the left and a plausible baseline failure on the right. Section~\ref{sec:quantitative-result} tests these distinctions across repeated trials.

\paragraph{Workspace coverage.}
The first prompt moves the editable copy of ``Unclear ownership delays handoffs'' from the offscreen archive into the report's Key slot. Three copies share the same text and appearance. Only the middle note is the editable copy. \method{} supplies each copy's identifier, lock state, and offscreen bounds in action coordinates. The baseline receives only coarse, unlabeled geometry for the archive.

\paragraph{Occluded object state.}
The second prompt moves the blue ``Handoff breakdown'' chart from a stack of fully overlapping drafts into the Chart slot. The screenshot shows only the top chart, and baseline summaries omit the hidden layers' colors. \method{} links each identifier to color, position in the layer stack, and occlusion state, making the blue chart addressable.

\paragraph{Native workspace structure.}
The third prompt separates materials for two study setups. All four cards appear under ``Study setup A,'' but the application records only the two Calls cards as children of that frame. The two Diaries cards are attached directly to the page and describe Setup B. \method{} exposes native parent identifiers so the agent can move the Diaries cards to Setup B while leaving Calls cards in Setup A.

\paragraph{Explicit connector relations.}
The final prompt moves the ``Lost decision context'' finding supported by Interview evidence into the Claim slot. Both candidates appear connected, but only one arrow is natively bound to the evidence and finding. The other merely touches them. \method{} exposes endpoint bindings to identify the supported finding and preserve its evidence connection during the move.

\section{Quantitative Result}
\label{sec:quantitative-result}

We test whether \method{} helps an agent identify the intended objects when screenshots and shape summaries do not expose the needed information. The four prompt families introduced in the Case Study (Figure~\ref{fig:case-study}) cover the editable offscreen copy, occluded chart color, native frame membership, and native connector binding. We created ten configurations. Three archive configurations each designate a different note as the editable copy. Three color configurations each make a different occluded chart blue. In two membership configurations, the Calls and Diaries pairs exchange frame membership. In two connector configurations, the evidence arrow binds to a different finding. Within each family, positions, labels, and the prompt remained fixed, so no rule based on position or identifier works across configurations.

Every experimental unit was a pair of trials. Both trials started with the same serialized workspace, camera, user prompt, empty conversation, language model, actions, executor, and review process. One received the baseline context. The other additionally received \method{}'s structured observation and its interpretation guide. Variant order within each pair and pair order were randomized. We repeated each of the ten configurations twenty times with each of three models, yielding $10\!\times\!20\!\times\!3=600$ pairs, or 1,200 individual trials. All trials used \texttt{claude-sonnet-4-5}, \texttt{gpt-5.4}, and \texttt{gpt-5.4-mini}, with temperature zero, optional reasoning or thinking disabled, and a deadline of 180 seconds. Transport errors and timeouts count as failures.

The automated evaluator compares the workspace state before and after each trial. A candidate counts as moved when its page position changes. \emph{Correct target} succeeds when the agent moves all and only the intended candidate objects, regardless of where it places them. \emph{Correct placement} is evaluated only among trials satisfying \emph{Correct target}. It succeeds when the moved objects preserve their size and other state, avoid overlapping one another by more than one square page unit, and lie fully inside the destination with an inset of 8 units. \emph{Full task success} requires \emph{Correct target} and \emph{Correct placement}, preservation of the expected native relation, and no additions, deletions, or updates to shapes outside the target set except the bound arrow that should follow its target. These outcomes distinguish choosing the right objects, placing them correctly, and completing the task without unintended changes.

\begin{table}[!t]
    \caption{Final results across all scheduled trials.}
    \label{tab:quantitative-result}
    \centering
    \footnotesize
    \setlength{\tabcolsep}{2.5pt}
    \begin{tabular*}{\columnwidth}{@{\extracolsep{\fill}}llccc@{}}
        \toprule
        Model & Variant & \shortstack{Correct\\target} & \shortstack{Correct\\placement} & \shortstack{Full task\\success} \\
        \midrule
        Sonnet 4.5 & \method{} & 94.5\% & 52.9\% & 49.5\% \\
        Sonnet 4.5 & Baseline  & 29.5\% & 32.2\% & 9.5\% \\
        GPT-5.4 & \method{} & 36.5\% & 68.5\% & 24.0\% \\
        GPT-5.4 & Baseline & 13.5\% & 29.6\% & 3.5\% \\
        GPT-5.4-mini & \method{} & 14.0\% & 10.7\% & 1.5\% \\
        GPT-5.4-mini & Baseline & 6.5\% & 0.0\% & 0.0\% \\
        \midrule
        Combined & \method{} & 48.3\% & 52.8\% & 25.0\% \\
        Combined & Baseline & 16.5\% & 27.3\% & 4.3\% \\
        \bottomrule
    \end{tabular*}
\end{table}

In the combined results for these ten fixed configurations and three models, \method{} performs better than the baseline on both \emph{Correct target} and \emph{Full task success}. It raises the \emph{Correct target} rate from 16.5\% to 48.3\% (paired difference 31.83 points, 95\% confidence interval (CI) [29.00, 34.67]) and the \emph{Full task success} rate from 4.3\% to 25.0\% (20.67 points, [17.67, 23.67]). When combined across models, the advantage appears in every prompt family. Gains in \emph{Correct target} range from 14.17 to 45.83 points, and the 95\% interval for every family excludes zero. Its size, however, depends on the model. Differences in \emph{Correct target} are 65.00 points for Sonnet~4.5 ([60.50, 69.50]), 23.00 for GPT-5.4 ([17.49, 29.00]), and 7.50 for GPT-5.4-mini ([3.00, 12.00]). Differences in \emph{Full task success} are 40.00, 20.50, and 1.50 points, respectively. The interval for GPT-5.4-mini includes zero ([0.00, 3.50]).

\emph{Correct placement} remains a distinct bottleneck. Of the 290 \method{} trials satisfying \emph{Correct target}, 153 (52.8\%) also satisfy \emph{Correct placement}. The corresponding baseline rate is 27/99 (27.3\%). These conditional rates are diagnostic rather than a matched causal comparison, but they show that satisfying \emph{Correct target} does not guarantee \emph{Correct placement}. One \method{} trial ended in a network error, while the baseline had seven timeouts at 180 seconds. All count as failures under the frozen protocol. Intervals use 10,000 paired bootstrap resamples stratified by model and configuration. They quantify stochastic outcomes from service calls for these ten fixed configurations, not generalization to a broader task population.

\paragraph{Operational measurements.}
Across the 600 initial \method{} requests, the model-facing JSON reconstructed from saved observations using the frozen serializer averaged 75.29~kB (range 73.51--77.05~kB; 1~kB = 1,000~bytes). Including a net 1.46~kB of interpretation guidance, additional text averaged 76.75~kB, excluding screenshots and shared baseline context. Table~\ref{tab:trial-duration} reports median durations for all scheduled trials (200 per model and condition), including failures and zero-action responses. Durations include browser preparation, agent execution, state export, and screenshot capture, so they do not isolate observation-generation or model-inference time.

\begin{table}[!t]
    \caption{Median trial duration (seconds).}
    \label{tab:trial-duration}
    \centering
    \normalsize
    \setlength{\tabcolsep}{3pt}
    \renewcommand{\arraystretch}{1.05}
    \begin{tabular}{|>{\raggedright\arraybackslash}p{\dimexpr0.44\columnwidth-2\tabcolsep-2\arrayrulewidth\relax}|>{\centering\arraybackslash}p{\dimexpr0.28\columnwidth-2\tabcolsep-\arrayrulewidth\relax}|>{\centering\arraybackslash}p{\dimexpr0.28\columnwidth-2\tabcolsep-\arrayrulewidth\relax}|}
        \hline
        \textbf{Model} & \textbf{Baseline} & \textbf{\method{}} \\
        \hline
        Sonnet 4.5 & 13.87 & 14.67 \\
        \hline
        GPT-5.4 & 2.17 & 2.22 \\
        \hline
        GPT-5.4-mini & 2.12 & 2.18 \\
        \hline
    \end{tabular}
    \par\vspace{-12pt}
\end{table}

\section{Discussion}

\paragraph{Observation as a design boundary.}
\method{} separates observation from action, allowing structured document state to supplement visual context while the editing pipeline remains fixed.

\paragraph{Correct target and Correct placement.}
Choosing the right objects does not ensure a correct edit. Even trials satisfying \emph{Correct target} can fail \emph{Correct placement} through coordinate interpretation, state preservation, collision avoidance, or containment. These outcomes should be evaluated separately. Relative placement primitives, actions that enforce constraints, and checks after each action may address placement failures but extend beyond observation.

\paragraph{Implications for visual workspaces.}
Visual workspaces can expose a semantic interface instead of requiring agents to reconstruct hidden document structure from pixels. Our design suggests preserving editable identity across visible and offscreen regions, distinguishing native relations from geometric inference, and expressing actionable geometry in the action API's coordinate frame. These principles may also apply to diagram editors, dashboard builders, and slide tools.

\paragraph{Limitations.}
Our evaluation covers ten targeted hidden-state configurations, limiting generalization. The four components and interpretation guidance were tested together, so their individual effects and the effects of representation versus information coverage remain unresolved. We did not compare against direct access to the native tldraw document. \method{} observes only the workspace state at prompt time. This spatial snapshot cannot reveal how objects and relations evolved, so an agent cannot distinguish established structure from recent or temporary arrangements. A temporal observation could summarize changes to show how the current layout developed.

\section{Conclusion}

We presented \method{}, a structured observation layer for agentic editing on 2D infinite canvases. Our evaluation across ten hidden-state configurations showed improved object selection while keeping the editing pipeline fixed. The results highlight the value of connecting visual context with editable object identities, document relations, and offscreen geometry.

\clearpage
\bibliographystyle{abbrv-doi}
\bibliography{references}

\end{document}